\documentclass[usenatbib,preprint]{mn2e}
\usepackage{graphicx}
\usepackage {natbib,aas_macros}
\usepackage {bm}
\usepackage {amsmath,amssymb}
\newcommand{\msun}{\thinspace M_\odot} 
\newcommand{\cm}{~{\rm g~cm}^{-3} }

\title[evolution of planetary-mass clumps]{Formation, orbital and  thermal evolution, and survival of planetary-mass clumps in the early
phase of circumstellar disk evolution}

\author[Tsukamoto et al]{Yusuke Tsukamoto$^{1}$, Masahiro
N. Machida$^{2}$ and  Shu-ichiro Inutsuka$^{1}$ \\
$^1$Department of Physics, Nagoya University, Furo-cho, Chikusa-ku, Nagoya, Aichi, Japan  \\
$^2$Department of Earth and Planetary Sciences, Kyushu University, 
6-10-1 Hakozaki, Higashi-ku, Fukuoka, Fukuoka, Japan \\
}

\begin{document}
\maketitle

\begin{abstract}
We report the results of our three-dimensional radiation hydrodynamics 
simulation of collapsing  unmagnetized
molecular cloud cores. We investigate the formation and evolution of the
circumstellar disk and the clumps formed by disk fragmentation.  
Our simulation shows that disk fragmentation occurs in the early
phase of circumstellar disk evolution and many clumps form.
The clump can be represented by a polytrope sphere
of index $n \sim 3$ and $n \gtrsim 4$ at central temperature 
$T_c \lesssim100$ K and $T_c \gtrsim 100$ K, respectively.
We demonstrate, numerically and theoretically, that the maximum mass 
of the clump, beyond which it inevitably collapses, 
is $\sim 0.03~M_{\odot}$. The entropy of the clump
increases during its evolution, implying that evolution is chiefly
determined by mass accretion from the disk rather than 
by radiative cooling.
Although most of the clumps rapidly migrate inward and finally fall
onto the  protostar, a few clumps remain in the disk.
The central density and temperature of the surviving 
clump rapidly increase and the clump undergoes a second collapse 
within $1000$ -- $2000$ years after its formation. 
In our simulation, three second cores of masses 
$0.2\msun$, $0.15\msun$, and $0.06\msun$ formed. These are
protostars or brown dwarfs rather than protoplanets.
For the clumps to survive as planetary-mass objects, 
the rapid mass accretion should be prevented by some mechanisms.
\end{abstract}


\section{Introduction}
Stars form in molecular cloud cores.
When the  angular momentum in the core is non-negligible, circumstellar disk formation is inevitable
because most of the gas cannot directly fall onto central protostar.
According to theoretical studies on the gravitational collapse of unmagnetized
molecular cloud cores, the protostar is surrounded by
a circumstellar disk immediately after its formation
\citep{1998ApJ...508L..95B,2010ApJ...724.1006M,2011MNRAS.416..591T,
2011MNRAS.417.2036B}. 

As noted by \citet{2010ApJ...718L..58I}, the resulting disk should
be sufficiently massive to develop gravitational instability (GI). 
If Toomre's $Q$ value of the disk is $\lesssim 1.5$, the disk is gravitationally unstable
to non-axisymmetric perturbation and develops spiral arms
\citep{1994ApJ...436..335L}. These spiral arms readjust the surface 
density \citep{2013ApJ...770...71T} 
and raise the disk temperature, thereby re-stabilizing the disk.
However, if radiative cooling is effective enough or the mass
accretion onto the disk is sufficiently high, the disk may
fragment and form clumps
\citep{2001ApJ...553..174G,2005MNRAS.364L..56R,
2010ApJ...718L..58I,2011ApJ...730...32S,2012PASJ...64..116K}.

%

Disk fragmentation is a candidate mechanism of wide-orbit 
planet formation 
\citep{2005MNRAS.364L..56R,2010ApJ...714L.133V,2011ApJ...729...42M,2013ApJ...768..131V}.
A wide-orbit planet is a planet separated from the central star by more than 10 AU \citep{2008Sci...322.1348M,2009ApJ...707L.123T,2009A&A...493L..21L,2010Natur.468.1080M,2010ApJ...719..497L}.
On the other hand, it has been suggested that disk fragmentation can also explain the formation of brown dwarfs 
\citep{2009MNRAS.392..413S,2011MNRAS.413.1787S} or multiple stellar systems
\citep{2008ApJ...677..327M,2010ApJ...708.1585K}.

The ultimate fate of the clumps depends upon their
orbital and internal evolution.
If the migration timescale is very small, 
the clump accretes onto the central protostar and eventually disappears.
On the other hand, if the collapse timescale of the clump is sufficiently long,
dust sedimentation may cause planetary embryos 
to form inside the clump \citep{2010MNRAS.408L..36N}.

Although the orbital and internal evolution of clumps in circumstellar disk is clearly important, 
a limited number of studies exist on the topic.
\citet{2011MNRAS.416.1971B} investigated the orbital evolution of 
massive planets formed by disk fragmentation. 
They showed that the planets rapidly migrate inward 
on a type I migration timescale \citep{2002ApJ...565.1257T}.
Adopting an analytical approach, \citet{2010MNRAS.408.2381N} showed 
that the collapse timescale of the
clumps is $\gtrsim 10^4$ years, considerably longer 
than the orbital timescale at $r \sim 100$ AU (approximately one thousand years).
However, Nayakshin used a simplified opacity model, and he ignored 
further mass accretion onto
the clumps from the disk.
Recently, \citet{2012MNRAS.427.1725G} investigated the internal
evolution of clumps which were extracted from three-dimensional (3D) global disc simulations.
They reported a collapse timescale of 
about several thousand years, shorter than the estimates of 
\citet{2010MNRAS.408.2381N}. However, these authors similarly neglected further mass
accretion onto the clumps.

To investigate the internal evolution of clumps permitting realistic gas accretion
from the disk, we must simultaneously calculate the evolution of both disk and 
clumps. 
Furthermore, appropriate treatment of 
radiative transfer and a realistic equation of state are crucial 
in studies of both the disk fragmentation \citep{2007ApJ...656L..89B} and internal
evolution of the clumps. 
Since clump evolution cannot be modeled assuming the
thin disk approximation, 3D simulations are required.
Two-dimensional simulations of the circumstellar disk would also
overestimate the extent of disk fragmentation (T. Tsuribe, private communication).
\citet{2009MNRAS.400.1563S}, who investigated the internal evolution of
clumps in 3D radiative SPH simulations,
reported  that clumps collapse on a timescale of $10^3 - 10^4$ years. 
However, they initially assumed a massive isolated
disk and it is unclear whether such a disk can be realized during the star formation process.

To realize a self-consistent study of disk fragmentation and 
clump evolution,  we conducted a 3D
radiation hydrodynamics simulation initiated
from gravitational collapse of a molecular cloud
core. Using this approach, we can follow the formation of a central
protostar, disk and its fragmentation. 
We can also follow the orbital and internal evolution of clumps with a
realistic mass accretion from the disk onto them.

In this study, we ignore the magnetic field and focus on the effects of
radiative cooling on the evolution of circumstellar disk and the internal
structure of clumps. Note however, that magnetic field may play an important
role in the formation and evolution of the circumstellar disk, because it can 
efficiently transfer angular momentum \citep[][]{2008ApJ...681.1356M,2009A&A...506L..29H,
2010ApJ...718L..58I,2011ApJ...738..180L, 2011ApJ...729...42M,2012A&A...543A.128J,2013MNRAS.432.3320S}.
We will describe the effect of magnetic field 
in our subsequent paper using new numerical methods for smoothed particle magnetohydrodynamics (SPMHD)
\citep[][]{2011MNRAS.418.1668I,2013MNRAS.434.2593T}.

This paper is organized as follows. In \S 2, we summarize the protostar 
formation process in a molecular cloud core 
and introduce the relevant terminology. The numerical method and initial condition are
described in \S 3, while \S 4, presents the results.  The paper concludes with a discussion 
in \S 5.

\begin{figure}
\includegraphics[width=100mm]{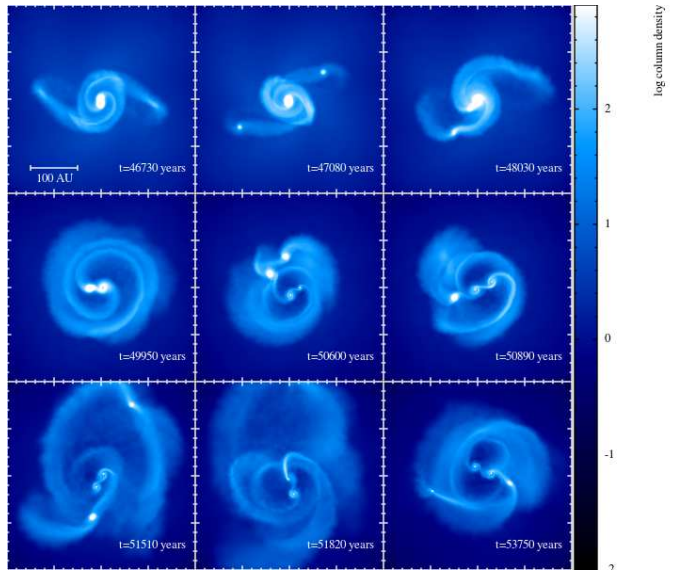}
\caption{
Time sequence of the surface density at the cloud center, viewed face-on.
The elapsed time after the cloud core begins to collapse is shown in each panel. 
}
\label{faceon_sigma}
\end{figure}

\begin{figure}
\includegraphics[width=100mm]{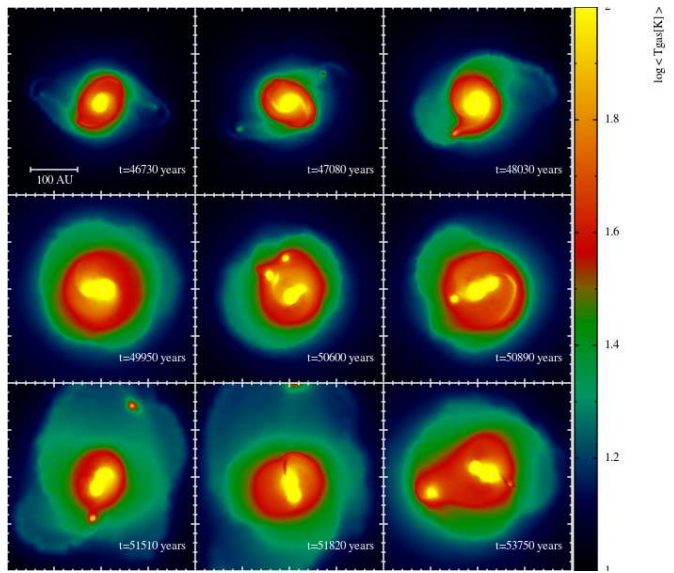}
\caption{
Same as in Fig.\ref{faceon_sigma}, but the gas temperature is plotted.
}
\label{faceon_Tgas}
\end{figure}

\section{From cloud core to protostar}
For later convenience, we summarize  protostar formation process in the spherically symmetric 
cloud core and introduce the relevant terminology.
The property of clumps formed in the disk are similar to the first core 
which is hydrostatic core formed during star formation. 
Thus, the internal evolution of the clump 
can be suitably described using 
the terminology of star formation  \citep[for more complete descriptions of the  protostar formation 
process, see ] [and references therein]
{1969MNRAS.145..271L,2000ApJ...531..350M,2012PTEP...770...71T}.

When the gravitational energy dominates the thermal energy around the center of the core, 
the cloud core begins to contract and the central density increases.
While the radiative cooling due to dust thermal emissions overwhelms the compressional heating,
the gas evolves isothermally, maintaining the temperature at about 10\,K (isothermal collapse phase).
The inner dense region collapses faster than the envelope because
the timescale of gravitational collapse is a decreasing function of density, causing runaway 
gravitational collapse.

At sufficiently high density ($\rho_{{\rm ad}} \sim 10^{-13} \cm$), 
the compressional heating catches up with radiative cooling and 
the isothermal collapse terminates.
The gas evolves almost adiabatically within the density range
 $10^{-13}{\rm g ~cm^{-3}} \lesssim \rho \lesssim 10^{-8}{\rm g ~cm^{-3}}$.
The gas temperature rises under adiabatic contraction with the adiabatic index, 
$\gamma=5/3$ for $T \lesssim100$\,K and  with $\gamma=7/5$ for $T\gtrsim100$\,K.
During this phase, thermal energy dominates over gravitational energy 
and a quasi-hydrostatic core forms. We refer to this quasi-hydrostatic core 
as the {\it first core}.

Once the central temperature of the first core reaches $\sim1500$\,K, 
the hydrogen molecules begin to dissociate. This endothermic reaction 
promotes a second round of gravitational collapse at
$10^{-8}{\rm g ~cm^{-3}} < \rho < 10^{-3}{\rm g ~cm^{-3}}$, known as the  {\it second collapse}.
Finally, when the molecular hydrogen is completely dissociated,
the gas again evolves adiabatically with $\gamma=5/3$ to form another hydrostatic core 
called as the {\it second core}.
The initial mass of and radius of the second core are  $M\sim 10^{-3} M_\odot$ 
and $r\sim 1 R_\odot$, respectively.
Only a small proportion of 
the first core collapses into the second core. Therefore,
the mass of the second core is rapidly increased by mass accretion from the remnant 
of the first core.

\section{Numerical Method and Initial Condition}
\label{method}
In this study, we extend the simulation code used in our previous studies 
\citep{2011MNRAS.416..591T,2013MNRAS.428.1321T}, to include radiative transfer
with flux-limited diffusion approximation according to
\citet{2004MNRAS.353.1078W, 2005MNRAS.364.1367W}.
Unlike these works, we adopted standard explicit scheme 
for the gas pressure and the artificial viscosity in the gas energy equation.
We adopted the equation of state (EOS) used in
\citet{2013ApJ...763....6T}, which involves seven species: 
${\rm H_2,~H,~H^+,~He,~He^+,He^{++}, e^-}$.
Molecular hydrogen is assumed as a 3:1 mixture of ortho- and para- hydrogen and
the translational, rotational and vibrational degrees of freedom are taken into account.
The hydrogen and helium mass fractions are $X=0.7$ and $Y=0.28$, respectively.

We used dust opacity table provided by \citet{2003A&A...410..611S} 
and gas opacity table by \citet{2005ApJ...623..585F}.
We did not use individual time-step technique for this work
and all particles were updated simultaneously.
When the density exceeds the threshold 
density, $\rho_{{\rm thr}} ~(5 \times 10^{-8} ~{\rm g\,cm^{-3}})$,
a sink particle was introduced.
Around $\rho_{\rm thr}$, the gas temperature reaches the dissociation 
temperature of molecular hydrogen ($T\sim 1500$\,K) and the second 
collapse begins in the clump. Therefore, we can follow the thermal evolution of the 
clump just prior to  second collapse.
The sink radius was set as $r_{{\rm sink}}=2$ AU.

The initial condition is an isothermal cloud core of uniform density, 
rigidly rotating with angular velocity 
$\Omega_0 = 1.4 \times 10^{-13} $\,s$^{-1}$. The mass, radius and temperature of the core 
are $M=1 M_\odot$, $r=3933$ AU and $T=10$\,K, respectively.
The resultant density is $\rho_0=2.3 \times 10 ^{-18} $\,g\,cm$^{-3}$.
The initial condition was subjected to a density perturbation, 
$\delta \rho$ $( =0.01\times \cos2\phi) $.
The ratios of thermal to gravitational energy $\alpha_0$ ($\equiv E_{\rm t}/E_{\rm g}$) 
and rotational to gravitational energy $\beta_0$ ($\equiv E_{\rm r}/E_{\rm g}$) 
are $\alpha_0=0.384$ and $\beta_0 = 0.01$, respectively, where $E_{\rm t}$, $E_{\rm r}$ 
and $E_{\rm g}$ denote the thermal, rotational and gravitational energy of
the initial cloud core.
These values are consistent with the results of recent 3D MHD simulation 
of molecular cloud and involved core formation \citep{2012ApJ...759...35I}.
The cloud core was modeled with about 530 000 SPH particles.

\section{Results}
\label{results}

\subsection{Overview of evolution}
Starting from a prestellar core,
we calculated the disk and clump evolutions
to $\sim 6000$\, years following the first fragmentation.
Figure~\ref{faceon_sigma} shows the surface density evolution at the center of the cloud core.
This figure indicates that many clumps can form in the circumstellar
disk during the early stages of disk evolution.
Because the clumps connect smoothly to the disk, they are not always clearly delineated.
In this paper, we defined
a clump as a gaseous object whose central 
density, $\rho_c$ is  $10^{-11}~{\rm g\,cm^{-3}} < \rho_c < \rho_{{\rm thr}} = 5 \times 10^{-8} \cm$.

When the central density exceeds  $10^{-11} \cm$, 
the clumps are clearly distinguished from the background gas (of typical
density, $\rho\lesssim 10^{-12} \cm$). At $\rho_c \sim \rho_{{\rm thr}}$, 
the central temperature exceeds the dissociation temperature of the hydrogen molecule 
($T\sim 1500$ K) and the second collapse begins.
We also defined an epoch of clump formation as the time when its central density
exceeds $10^{-11}~{\rm g\,cm^{-3}}$, although gravitational contraction
begins at lower density. 
Since the central first core is not formed in the disk, its thermal evolution is
different from that of the other clumps.
Therefore, we do not regard it as a clump.
The central density of the first core exceeded $\rho_{{\rm thr}}$
at $t=47500$ years (between Fig.~\ref{faceon_sigma}{\it b} 
and Fig.~\ref{faceon_sigma}{\it c}) and the first sink particle was introduced at this stage.

 Two clumps formed (Fig.~\ref{faceon_sigma}{\it b}) by the disk fragmentation.
They rapidly migrated to the center and accreted onto the sink particle
(Fig.~\ref{faceon_sigma}{\it c}). 
During this migration, a new clump formed at $r \sim 80-100$\,AU 
(Fig.~\ref{faceon_sigma}{\it c}) and similarly migrated rapidly to the center. 
As it journeyed, the new clump gathered mass from the disk, increasing its
 central density and temperature
(Fig.~\ref{faceon_sigma}{\it d}).
The central density of the clump 
exceeded $\rho_{{\rm thr}}$ and  the second sink particle was introduced 
between Fig.~\ref{faceon_sigma}{\it d} and {\it e}. 
The two sink particles were gravitationally bound 
and formed a binary.
As seen in Fig.~\ref{faceon_sigma}{\it e}-{\it g}, 
clump formation continued in the circumbinary disk
throughout the simulation time, despite the presence of the second sink
(Fig.~\ref{faceon_sigma}{\it e}-{\it g}). 
Most of the clumps rapidly accreted onto the sink particles and disappeared.
However, one clump formed relatively far from both sink particles
(see upper region of Fig.~\ref{faceon_sigma}{\it g}) survived.
The central density of this clump exceeded $\rho_{{\rm thr}}$ and a third
sink particle was introduced.

Seven clumps formed within 6000\,years after the first clump formation, 
of which five migrated to the central region and finally 
accreted onto the sink particles.
The two clumps  
survived to second collapse.
At the end of the calculation, the masses of the sink particles (in order of decreasing age) 
were $0.2\msun$, $0.15\msun$, and $0.06\msun$, respectively.

Figure~\ref{faceon_Tgas} shows the gas temperature at the 
epochs indicated in Fig.~\ref{faceon_sigma}. Collectively, Figs.~\ref{faceon_sigma} 
and \ref{faceon_Tgas} reveal a weak  
correlation between density and temperature.  
The gas temperature is rather determined by the gravitational potential.
Therefore, EOSs that assume a polytropic relationship between 
pressure and  density, $p=p(\rho)$ (or the barotropic approximation) 
do not appropriately describe the thermal evolution of the disk.
We also observe that clumps form in the outer cold regions of the disk, 
where the gas temperature is several tens Kelvin.
Once a clump has formed, its temperature rapidly increases to higher than $100$ K under
compressional heating.

\begin{figure}
\includegraphics[width=70mm,angle=-90]{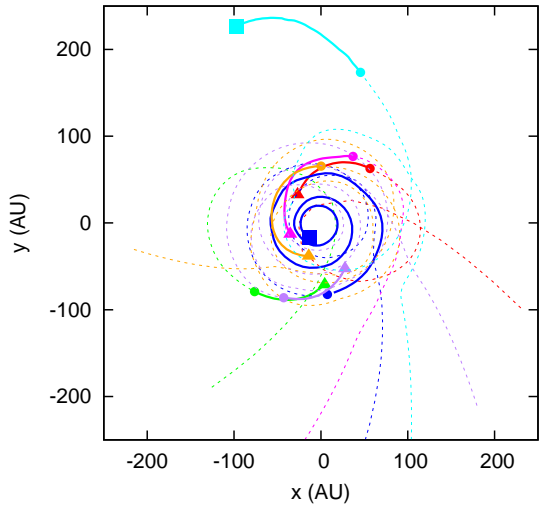}
\caption{
The orbits of the representative gas particles are shown.
Dashed and solid lines represent the orbits before and after the clump 
formation ($\rho_c>10^{-11} \cm$), respectively.
Symbols mark the positions where the clumps forms (circles), where the clump central
density begins to decrease (triangles), and where the second collapse begins in the clump (or the 
sink particles are inserted; square).
}
\label{orbit}
\end{figure}

\begin{figure}
\includegraphics[width=60mm,angle=-90]{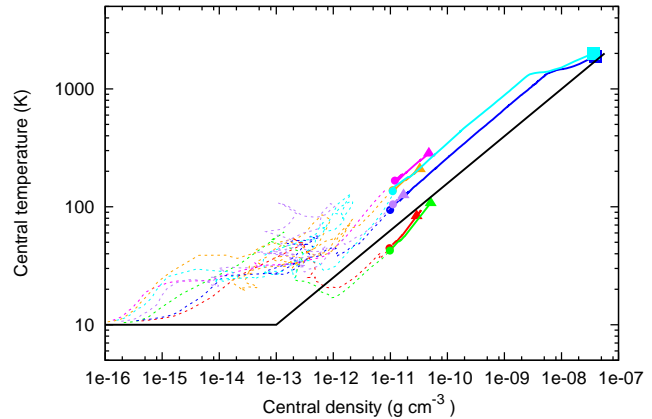}
\caption{
Temperature of representative gas particles as a function of 
density. Symbols are explained in the caption of Fig.~\ref{orbit}.
The clump that is identical in Figs. \ref{orbit} and \ref{rho_T} is indicated by the same color.
The black solid line plots the typical evolution of the barotropic approximation.
}
\label{rho_T}
\end{figure}

\begin{figure}

\includegraphics[width=60mm,angle=-90]{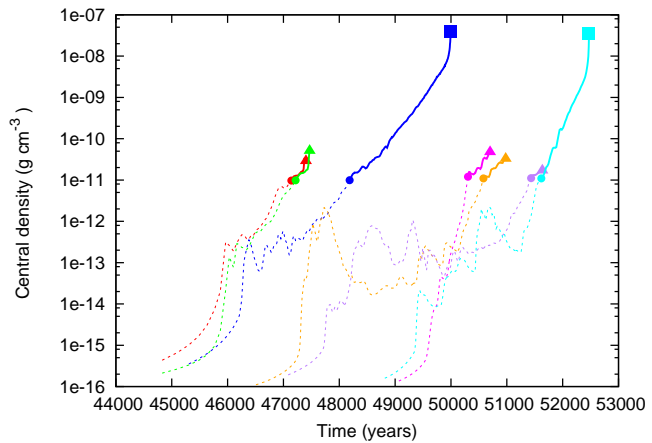}
\caption{
Density of representative gas particles as a function of elapsed time.
Symbols are explained in the caption of Fig.~\ref{orbit}.
The clump that is identical in Figs. \ref{orbit} and \ref{time_rho} is indicated by the same color.
}
\label{time_rho}
\end{figure}

\subsection{Orbital and internal  evolution of clumps}
In this subsection, we investigate the orbital and thermal
evolution of the clumps.
We also investigate the orbital and thermal history of the gas prior to clump formation by
tracking the fluid element at the
center of each clump.
To this end, we identify {\it representative gas particles} 
that reside at the center of each clump.
When a clump undergoes a second collapse,
the representative gas particle is defined as the gas particle which has maximum density immediately 
before the sink particle is introduced.
If a clump is tidally disrupted and accreted onto the
sink particle, its central density begins to decrease at
some epoch. In this case, the representative gas particle is defined as
the particle which has the maximum density in the clump, at the time 
when the central density of clump is maximum.
With this definition, 
the representative gas particle traces the evolution of the center of each 
clump.

Figure~\ref{orbit} shows the orbits of the representative gas particles. 
In this figure, the orbits of formed clumps (particle densities exceeding $10^{-11} \cm$), 
are plotted with solid lines,
while the preceding orbits are plotted with dashed lines. 
According to Fig. \ref{orbit}, the gas particles orbit within the disk 
for several orbital periods after they have accreted onto the disk. 
The ensuing gravitational contraction leads to clump formation.
Most of the clumps are destroyed by tidal disruption and accrete 
onto the sink particle(s) in less than one orbital period.   
A notable exception is the clump shown by the cyan line. The gas
was kicked out by the gravitational interaction between
the binary and the clump formed relatively far from the
binary. This clump was granted sufficient time for second collapse.

Figure \ref{rho_T} shows the temperature evolution of the representative
gas particles as a function of density. 
In this figure,
a typical temperature evolution of the barotropic approximation is
shown for comparison \citep[see, for example,][]{2011MNRAS.417.2036B}.

Unlike the barotropic approximation, the gas temperature
increases to several tens Kelvin at $\rho=10^{-15}-10^{-14} \cm$. 
While the gas particles orbit in the disk, they undergo the complex
density and temperature evolution around  $\rho \sim10^{-14}-10^{-12}\cm$. 
The temperature of the gas can increase to $100$
K due to the heating caused by the gravitational instability. 
Once the density exceeds $\gtrsim 10^{-12} \cm$, 
the gas becomes adiabatic and further evolution of the
central temperature of the clumps consistent with adiabatic contraction. 
However, the complex 
thermal evolution at densities around $\rho \sim10^{-14}-10^{-12} \cm$ 
induces the large variance of the central entropies.
As described above, most of the clumps are tidally disrupted  
prior to the second collapse.
However, in two instances (clump evolution shown by blue and cyan lines 
in Figs. \ref{orbit} and \ref{rho_T}), the central temperature reaches
the dissociation temperature of the molecular hydrogen, and second collapse is initiated. 
Dissociation is revealed by shallower slope at $T\gtrsim 1500$ K in Fig. \ref{rho_T}.

The collapse timescales of the clumps (timescales required for second collapse) are 
of particular interest because some interesting planet formation scenarios assume the long collapse
timescales of the clumps \citep[e.g.,][]{2010MNRAS.408L..36N,2011MNRAS.415.3319C}. 
Long collapse timescales also enable direct observation of the clumps.
Figure \ref{time_rho} shows how the density of 
the representative gas particles evolves over 
time. When the gas accretes onto the disk, its density rapidly increases 
from $\rho \sim 10^{-16} $ to $ 10^{-13} \cm$ and 
oscillates between $\rho \sim 10^{-14} -
10^{-12} \cm$ within the disk. 
Gravitational contraction (occurring  at $\rho \gtrsim 10^{-12} \cm$) is accompanied 
by a rapid increase in density.
Although most of the clumps disappeared within several hundred years, two clumps survived and 
collapsed over a timescale of $1000$ -- $2000$ years (blue and cyan lines). 
This timescale is much shorter than that estimated by
\citet{2010MNRAS.408.2381N}. 

\subsection{Structure of clumps}
\begin{figure*}

\includegraphics[width=60mm,angle=-90]{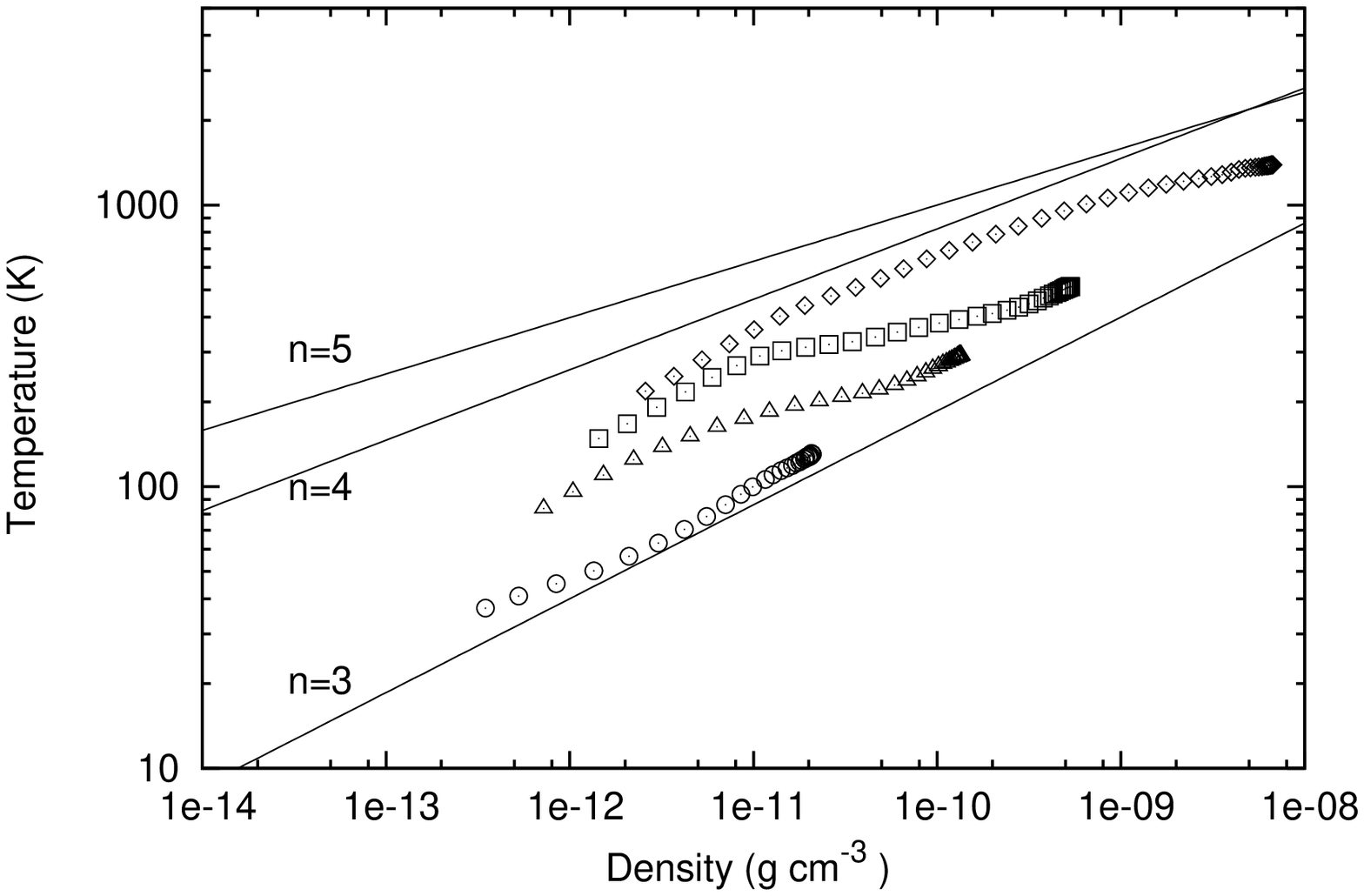}
\includegraphics[width=60mm,angle=-90]{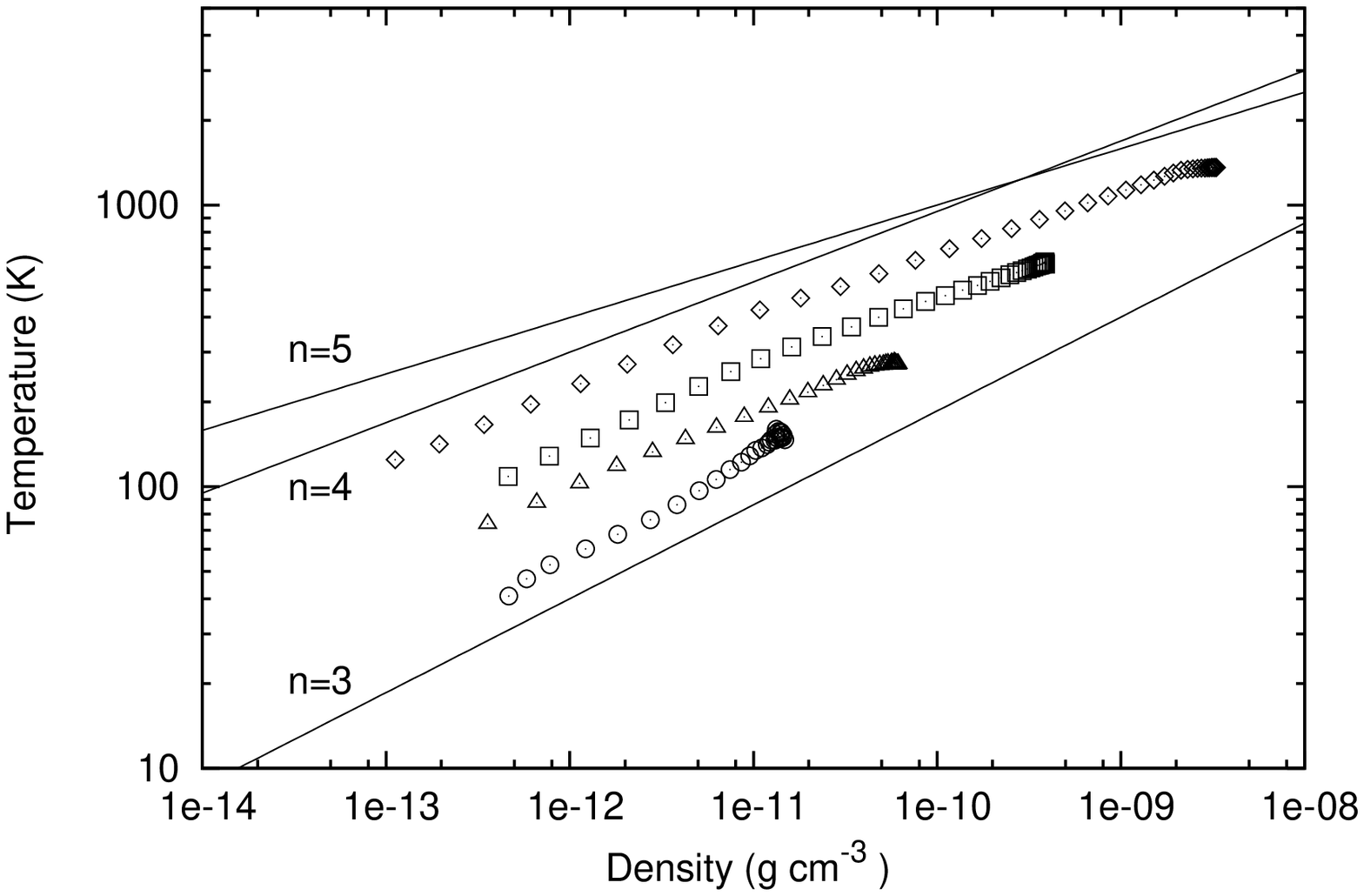}
\caption{
Profiles of clumps 1 (left) and 2 (right) in the
density-temperature plane within 10 AU of the clump center.
Symbols denote the profiles at different epochs.
The central density  of each epoch is  
$\rho_c= 2.1 \times 10^{-11}\cm$ (circles), ~$1.3 \times 10^{-10}\cm$ (triangles), 
~ $ 1.5 \times 10^{-9} \cm $ (rectangles) and ~$6.7 \times 10^{-9}\cm$ (rhombi)
 for clump 1 and $\rho_c= 1.6 \times 10^{-11}\cm$ (circles), ~$6.2 \times 10^{-11}\cm$ 
(triangles),~$3.9 \times 10^{-10}\cm$ (rectangles) and ~$3.3 \times 10^{-9}\cm$ (rhombi)
for clump 2.
Solid lines show the polytropic relation, $T \propto \rho^{\frac{1}{n}}$ ($n=3,~4,~5$). 
}
\label{rho_T_prof}
\end{figure*}

In figure \ref{rho_T_prof}, for the two
clumps that survived to second collapse, temperature is plotted as a function of 
density at different epochs (the clumps evolve
from the circles to the rhombi in the figure) to investigate 
how the clump structure evolves.
The profiles within 10 AU from the center of the clumps are shown.
The clump that collapses at $t=50 000$ years (blue
lines in figure \ref{orbit}, \ref{rho_T}, \ref{time_rho}) is referred 
as clump 1, while that collapsing at
$t=52500$ years  (cyan lines in figure \ref{orbit},
\ref{rho_T}, \ref{time_rho}) is denoted clump 2. 
The polytropic relationship, $T \propto \rho^{\frac{1}{n}}$ ($n=3,~4,~5$), 
where $\rho$ and $T$ are the density and 
temperature, respectively, is plotted for comparison. 

The clump structures are adequately modeled by 
the polytrope of index $n \sim 3$ at central temperatures, $T_c$ is $ \lesssim 100$
K. As the central temperature increases, the profiles become shallower, and
they can be represented with the polytropes of $n \gtrsim 4$ at
central temperatures exceeding $100  $ K.
However, the structure of clump 1 is distorted and a single-index
polytrope yields a poor fit at
$T_c>100  $ K.
On the other hand, the single-index polytrope sufficiently fits the structure of clump 2
at these temperature.

The structural difference between clumps 1 and 2 is attributable to 
the entropy of the accretion flow.
The rapid inward migration of clump 1 (see, Fig. 3) is accompanied by rapid
entropy changes of the ambient disk gas, which distort the clump structure.
On the other hand, the semi-major axis of clump 2 remains relatively constant 
suggesting that the entropy of the
accretion flow also changes little throughout the clump evolution.

As shown in Fig. 6, the
entropy of the clumps increases during their evolution. 
This is easily understood from the increased temperature
at fixed density as the clump evolves.
Entropy is introduced by mass accretion from the disk. 
Therefore,  mass accretion 
plays an important role in the structural evolution of the
clumps. In clumps evolving solely by
radiative cooling, the entropy declines over time \citep[see, for example, Chp. 17 of][]{1968Cox}.

\subsection{Mass evolution of clumps}
\begin{figure*}
\label{mass}
\includegraphics[width=60mm,angle=-90]{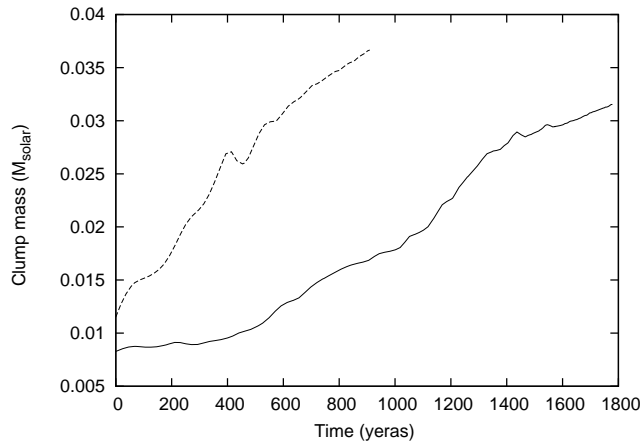}
\caption{
The  mass of the clump 1 (solid) and clump 2 (dashed) as a
 function of the elapsed time after the clump formation.}
\label{rad_prof}
\end{figure*}

The mass evolution of the clumps is plotted in figure \ref{rad_prof}. 
The solid and dashed lines denote the masses of clumps 1 and 2, respectively.
Because the clumps smoothly connect to the disk, their masses cannot be directly specified.
Here, we define the mass of the clump, $M_c$, so that it satisfies
\begin{eqnarray}
\label{def_mass}
3\int^{M_c}_{0} \frac{p}{\rho}dM_r=\int^{M_c}_{0}\frac{GM_r}{r}dM_r,
\end{eqnarray}
where $p,~\rho$ and $M_r$ signify the pressure, density, and cumulative mass at $r$, respectively. 
Note that equation (\ref{def_mass}) is identical to Virial theorem,
if the surface pressure is negligible and the clump is
in hydrostatic equilibrium.
The discretized form of equation (\ref{def_mass}) is,
\begin{eqnarray}
\label{disc_mass}
3\sum_j \frac{p_j}{\rho_j}m_j=\sum_j \frac{GM_{r<r_j}}{r_j}m_j,
\end{eqnarray}
where $p_j,~\rho_j,~m_j$ and $r_j$ are the pressure, density, mass, and radius from the 
clump center of the {\it j}-th particle, respectively.
Summation is performed in ascending order from the particle of smallest radius.
$M_{r<r_j}$ is the total mass within $r_j$. 
As shown in Fig. \ref{rad_prof}, the clumps continually aggregate mass throughout their
evolution. Clump 1 and 2 undergo second collapse at 900 years and 1800 years after its formation,
respectively.
The clump mass at second collapse is $\sim 0.03~ M_{\odot}$, consistent with the 
maximum clump mass predicted for a
polytrope sphere (see, Equation (\ref{critical_mass}) in \S 5). 

The mass accretion rate
onto the clump can be estimated from Fig. \ref{rad_prof}. 
The clump mass increases by approximately $ 0.01
M_{\odot}$ within $ 1000$ years, yielding a mass accretion rate
onto the clumps, 
$ \dot{M}_c \sim  10^{-5}  ~M_{\odot}~ {\rm years}^{-1}$. 
This result also accords with the theoretical
estimate (see, Equation (\ref{accretion_rate}) in \S 5).

\section{Summary and Discussion}
\label{discussion}
In this paper, we performed a radiation hydrodynamics simulation using a realistic equation of
state and investigated  the orbital and internal evolution of clumps formed by disk fragmentation.
Note that we investigated the formation and evolution of disk and clumps self-consistently.

According to our simulation, disk fragmentation and clump formation occur 
in the early evolution phase of the circumstellar disk.
Disk fragmentation is induced by mass accretion from the infalling 
envelope. These results are consistent with previous works
\citep{2011ApJ...729...42M,2012ApJ...746..110Z}.
We show that most of the clumps rapidly 
migrate inward and accrete onto protostar(s). 
However, some clumps may survive to second collapse.

The disk gas and clump centers show the following evolutionary trends in
their density and temperature.
At densities of $10^{-14} \cm \lesssim \rho \lesssim 10^{-12} \cm$, the gas resides in the disk, where it
 undergoes complex thermal evolution. 
Heating caused by gravitational instability can raise the gas temperature to 
 $T\sim 100$ K.  When the density exceeds $\sim 10^{-12} \cm$, gravitational contraction begins 
and clumps form.
The central gas of the clump evolves adiabatically and further evolution 
typifies that of first cores \citep{2000ApJ...531..350M}. 
However, reflecting the complex thermal history, 
the central entropy of the clumps has large deviation.
Once gravitational contraction has begun, the central density increases
rapidly toward second collapse within
$1000$ - $2000$ years. 
We found that the clumps are adequately modeled by the polytrope spheres
of index $n \sim 3$ at central temperatures $\lesssim100$ K and $n
\gtrsim 4$ at higher central temperature.
These indices are much higher than those used in \citet{2012ApJ...746..110Z}.

Fixing the polytropic index, the clump mass can be estimated as a 
function of the central temperature, $T_c$ and central density, $\rho_c$.
During adiabatic contraction, 
the entropy is constant and $T_c/\rho_c^{\gamma-1}=T_{{\rm ad}}/\rho_{{\rm ad}}^{\gamma-1}$,
assuming constant heat capacity at constant volume $c_v$ and constant ratio of heat capacity $\gamma$.
we define the adiabatic density, $\rho_{{\rm ad}}$
and adiabatic temperature, $T_{{\rm ad}}$ as the
values above which the central gas in the clump evolves adiabatically.
The mass of the clumps is then given as 
\begin{equation}
\begin{split}
\label{critical_mass}
M_c=&(n+1)^{3/2} \left[ \frac{k_{\rm B}^3}{4\pi G \mu^3m_{\rm u}^3}\frac{T_c^3}{\rho_c}\right]^{\frac{1}{2}} \left[- \xi^2 \frac{d \theta}{d \xi} \right]_{\xi=\xi_n} \\
=&\left\{\
     \begin{array}{c}
      3.37 \times 10^{-2}    
 ~(n=4) \\
      2.71 \times 10^{-2}    
 ~(n=3)
     \end{array}
     \right\} \\
 & \times  \left(\frac{T_c}{1000 {\rm K}}\right)^{\frac{1}{4}}\left(\frac{\rho_{{\rm ad}}}{10^{-14}
       \cm}\right)^{-\frac{1}{2}}\left(\frac{T_{{\rm ad}}}{10 {\rm K}}\right)^{\frac{5}{4}}
 M_{\odot},
\end{split}
\end{equation}
where, $k_{\rm B}$, $\mu$ and $m_{\rm u}$ are the Boltzmann constant, mean molecular weight, and atomic mass, respectively.
In this estimate, $c_v$ and $\gamma$ ($=7/5$) are assumed constant for simplicity,
and the clump mass is expressed in terms of central temperature, 
adiabatic density, and adiabatic temperature.
Molecular hydrogen
begins to dissociate at central temperatures $\sim 1000$ K, the temperature of
second collapse. 
Thus, equation~(\ref{critical_mass}) indicates that second collapse occurs 
when the clump mass reaches $\sim 0.03~M_{\odot}$.
In other words, the clump has a maximum mass and second collapse is inevitable in clumps of mass
 $\sim 0.03~M_{\odot}$.
Therefore, the timescale of clump collapse is related 
to the mass accretion rate.
This estimate of the maximum mass agrees favorably with our simulation results.
The two clumps formed in our simulation collapsed 
at  masses $0.036 \msun$ and $0.031 \msun$ (see Fig. 7).

We also showed that the clump entropy increases as the clump
evolves. This implies that clump evolution is driven by mass accretion.
By contrast, entropy decreases in clumps evolving by radiative
cooling alone.
The mass accretion rate onto the clump can be estimated by the 
Bondi-Hoyle-Lyttleton \citep{1944MNRAS.104..273B} accretion rate as
\begin{equation}
\begin{split}
\label{accretion_rate}
\dot{M_c}=& \frac{2 \pi G^2 M_c^2
 \rho_{\infty}}{(c_{\infty}^2+v_{\infty}^2)^{3/2}} \\
=& 2.7 \times 10^{-5} \\
  & \times \left(\frac{M_c}{0.01 {\rm M_{\odot}}}\right)^{2}\left(\frac{\rho_{\infty}}{10^{-14}
       \cm}\right)\left(\frac{\sqrt{c^2_\infty+v_{\infty}^2}}{400 ~{\rm m~s^{-1} } }\right)^{-3} M_{\odot}~{\rm year}^{-1},
\end{split}
\end{equation}
where, $\rho_\infty$, $c_\infty$, and  $v_\infty$
are  the density, sound velocity and the characteristic velocity of the gas far from the clump, respectively. 
Here, we assumed  the typical velocity of the gas as $400 ~{\rm m~s^{-1}}$, which is close 
to the value of the sound velocity at $T\sim 40$ K.
From the accretion rate, the timescale of
second collapse is estimated as $ 1000$ years 
($\simeq0.03M_{\odot}/2.7\times10^{-5}M_\odot$yr$^{-1}$), consistent with 
the collapse timescale of the simulated clumps 
(see, Fig. 7), but  much smaller than the estimates of \citet{2010MNRAS.408.2381N}. 
This difference is largely attributable to the negligence of 
 further mass accretion in the earlier study. Our simulation
suggests that mass accretion is a main driver of clump collapse.

The typical luminosity of clumps in circumstellar disks has been estimated 
as $L \sim 10^{-3} L_\odot$ \citep[see, for example, ][]
{2012ApJ...746..110Z}.
Here, we emphasize
that clump luminosity would 
significantly increase at second collapse,
as the clump gas rapidly accretes
onto the new born second core. The luminosity is estimated as
\begin{equation}
\begin{split}
L_{\rm{acc}} =& \frac{GM_{{\rm pr}}\dot{M}_{{\rm pr}}}{R_{{\rm pr}}}\\
 =&  25 \times \left(\frac{M_{{\rm pr}}}{0.01 {\rm M_{\odot}}}\right)
 \left(\frac{\dot{M}_{{\rm pr}}}{2.4 \times 10^{-4} {\rm M_{\odot}} {\rm year}^{-1}}\right)
 \left(\frac{R_{{\rm pr}}}{3 R_{\odot}}\right)^{-1} L_{\odot},
\end{split}
\end{equation}
where  $M_{{\rm pr}}$, $R_{{\rm pr}}$, and $\dot{M}_{{\rm pr}}$ are 
the second core mass,  second core radius
and the mass accretion rate onto the second core, respectively.
The mass accretion rate is estimated as
\begin{equation}
\dot{M}_{\rm pr}=\frac{c_s^3}{G}=2.4 \times 10^{-4}  \left(\frac{c_s}{1000 ~{\rm m~s^{-1}}} \right)^{3}~{\rm M_{\odot}}~{\rm year}^{-1}.
\end{equation}
$3 R_\odot$ is the typical radius of a young protostar \citep[e.g.,][]{1993ApJ...418..414P,2000ApJ...531..350M}.
This luminosity increase would continue over about $M_{c}/\dot{M}_{{\rm pr}} \sim 100$ years.

Following second collapse, the newly formed second core accumulates further mass from
the disk.
The final masses of the sink particles emerging from our simulation were $0.2\msun$,
$0.15\msun$, and $0.06\msun$ at the end of the calculation.
Thus, they are rather protostars or brown dwarfs than protoplanets. This is consistent with
the results of \citet{2009MNRAS.400.1563S,2009MNRAS.392..413S}.

In our simulation, the formed clumps either fall onto the 
protostar(s) and disappear or evolve into protostars or brown dwarfs.
The clump mass easily exceeds the planetary mass range ($\lesssim 0.01 \msun$). 
Therefore, if disk fragmentation is responsible for the wide-orbit planets 
found in recent observations
\citep[e.g.,][]{2010Natur.468.1080M}, 
the rapid migration to the central star should be
avoided, and the mass of the clumps should be kept small. 
Although the latter condition is little 
recognized, it is problematic in explaining 
how distant planets can emerge from disk fragmentation. 
To retain small mass clump and avoid rapid inward migration, 
an additional mechanism that decouples the disk and the clumps may be required.
We will investigate such mechanisms in a more realistic setup involving magnetic fields
\citep[see,][Iwasaki in preparation]{2011MNRAS.418.1668I,2013MNRAS.434.2593T}

\section *{Acknowledgments}
We thank  T. Tsuribe, K. Iwasaki, S. Okuzumi, E.I. Vorobyov, and K. Tomida for their fruitful
discussions.
We also thank K. Tomida and Y. Hori to provide their EOS table to us.
The snapshots were produced by SPLASH \citep{2007PASA...24..159P}.
The computations were performed on a parallel computer, XT4 system at CfCA of
NAOJ and SR16000 at YITP in Kyoto University.
Y.T. is financially supported by Research Fellowships of JSPS for Young Scientists.

\bibliography{article}

\begin{thebibliography}{53}
\expandafter\ifx\csname natexlab\endcsname\relax\def\natexlab#1{#1}\fi

\bibitem[{{Baruteau}, {Meru} \& {Paardekooper}(2011){Baruteau}, {Meru}, \&
  {Paardekooper}}]{2011MNRAS.416.1971B}
{Baruteau} C., {Meru} F., {Paardekooper} S.-J., 2011, \mnras, 416, 1971

\bibitem[{{Bate}(1998)}]{1998ApJ...508L..95B}
{Bate} M.~R., 1998, \apjl, 508, L95

\bibitem[{{Bate}(2011)}]{2011MNRAS.417.2036B}
---, 2011, \mnras, 417, 2036

\bibitem[{{Boley} {et~al}\mbox{.}(2007){Boley}, {Hartquist}, {Durisen}, \&
  {Michael}}]{2007ApJ...656L..89B}
{Boley} A.~C., {Hartquist} T.~W., {Durisen} R.~H., {Michael} S., 2007, \apjl,
  656, L89

\bibitem[{{Bondi} \& {Hoyle}(1944)}]{1944MNRAS.104..273B}
{Bondi} H., {Hoyle} F., 1944, \mnras, 104, 273

\bibitem[{{Cha} \& {Nayakshin}(2011)}]{2011MNRAS.415.3319C}
{Cha} S.-H., {Nayakshin} S., 2011, \mnras, 415, 3319

\bibitem[{{Cox} \& {Giuli}(1968)}]{1968Cox}
{Cox} J.~P., {Giuli} R.~T., 1968, Gordon and Breach, science publishers

\bibitem[{{Ferguson} {et~al}\mbox{.}(2005){Ferguson}, {Alexander}, {Allard},
  {Barman}, {Bodnarik}, {Hauschildt}, {Heffner-Wong}, \&
  {Tamanai}}]{2005ApJ...623..585F}
{Ferguson} J.~W., {Alexander} D.~R., {Allard} F., {Barman} T., {Bodnarik}
  J.~G., {Hauschildt} P.~H., {Heffner-Wong} A., {Tamanai} A., 2005, \apj, 623,
  585

\bibitem[{{Galvagni} {et~al}\mbox{.}(2012){Galvagni}, {Hayfield}, {Boley},
  {Mayer}, {Ro{\v s}kar}, \& {Saha}}]{2012MNRAS.427.1725G}
{Galvagni} M., {Hayfield} T., {Boley} A., {Mayer} L., {Ro{\v s}kar} R., {Saha}
  P., 2012, \mnras, 427, 1725

\bibitem[{{Gammie}(2001)}]{2001ApJ...553..174G}
{Gammie} C.~F., 2001, \apj, 553, 174

\bibitem[{{Hennebelle} \& {Ciardi}(2009)}]{2009A&A...506L..29H}
{Hennebelle} P., {Ciardi} A., 2009, \aap, 506, L29

\bibitem[{{Inoue} \& {Inutsuka}(2012)}]{2012ApJ...759...35I}
{Inoue} T., {Inutsuka} S.-i., 2012, \apj, 759, 35

\bibitem[{{Inutsuka}(2012)}]{2012PTEP...770...71T}
{Inutsuka} S., 2012, Prog. Theor. Exp. Phys., 2012, 307

\bibitem[{{Inutsuka}, {Machida} \& {Matsumoto}(2010){Inutsuka}, {Machida}, \&
  {Matsumoto}}]{2010ApJ...718L..58I}
{Inutsuka} S., {Machida} M.~N., {Matsumoto} T., 2010, \apjl, 718, L58

\bibitem[{{Iwasaki} \& {Inutsuka}(2011)}]{2011MNRAS.418.1668I}
{Iwasaki} K., {Inutsuka} S., 2011, \mnras, 418, 1668

\bibitem[{{Joos}, {Hennebelle} \& {Ciardi}(2012){Joos}, {Hennebelle}, \&
  {Ciardi}}]{2012A&A...543A.128J}
{Joos} M., {Hennebelle} P., {Ciardi} A., 2012, \aap, 543, A128

\bibitem[{{Kimura} \& {Tsuribe}(2012)}]{2012PASJ...64..116K}
{Kimura} S.~S., {Tsuribe} T., 2012, \pasj, 64, 116

\bibitem[{{Kratter} {et~al}\mbox{.}(2010){Kratter}, {Matzner}, {Krumholz}, \&
  {Klein}}]{2010ApJ...708.1585K}
{Kratter} K.~M., {Matzner} C.~D., {Krumholz} M.~R., {Klein} R.~I., 2010, \apj,
  708, 1585

\bibitem[{{Lafreni{\`e}re}, {Jayawardhana} \& {van
  Kerkwijk}(2010){Lafreni{\`e}re}, {Jayawardhana}, \& {van
  Kerkwijk}}]{2010ApJ...719..497L}
{Lafreni{\`e}re} D., {Jayawardhana} R., {van Kerkwijk} M.~H., 2010, \apj, 719,
  497

\bibitem[{{Lagrange} {et~al}\mbox{.}(2009){Lagrange}, {Gratadour}, {Chauvin},
  {Fusco}, {Ehrenreich}, {Mouillet}, {Rousset}, {Rouan}, {Allard}, {Gendron},
  {Charton}, {Mugnier}, {Rabou}, {Montri}, \& {Lacombe}}]{2009A&A...493L..21L}
{Lagrange} A.-M. {et~al.}, 2009, \aap, 493, L21

\bibitem[{{Larson}(1969)}]{1969MNRAS.145..271L}
{Larson} R.~B., 1969, \mnras, 145, 271

\bibitem[{{Laughlin} \& {Bodenheimer}(1994)}]{1994ApJ...436..335L}
{Laughlin} G., {Bodenheimer} P., 1994, \apj, 436, 335

\bibitem[{{Li}, {Krasnopolsky} \& {Shang}(2011){Li}, {Krasnopolsky}, \&
  {Shang}}]{2011ApJ...738..180L}
{Li} Z.-Y., {Krasnopolsky} R., {Shang} H., 2011, \apj, 738, 180

\bibitem[{{Machida}, {Inutsuka} \& {Matsumoto}(2010){Machida}, {Inutsuka}, \&
  {Matsumoto}}]{2010ApJ...724.1006M}
{Machida} M.~N., {Inutsuka} S., {Matsumoto} T., 2010, \apj, 724, 1006

\bibitem[{{Machida}, {Inutsuka} \& {Matsumoto}(2011){Machida}, {Inutsuka}, \&
  {Matsumoto}}]{2011ApJ...729...42M}
---, 2011, \apj, 729, 42

\bibitem[{{Machida} {et~al}\mbox{.}(2008){Machida}, {Tomisaka}, {Matsumoto}, \&
  {Inutsuka}}]{2008ApJ...677..327M}
{Machida} M.~N., {Tomisaka} K., {Matsumoto} T., {Inutsuka} S., 2008, \apj, 677,
  327

\bibitem[{{Marois} {et~al}\mbox{.}(2008){Marois}, {Macintosh}, {Barman},
  {Zuckerman}, {Song}, {Patience}, {Lafreni{\`e}re}, \&
  {Doyon}}]{2008Sci...322.1348M}
{Marois} C., {Macintosh} B., {Barman} T., {Zuckerman} B., {Song} I., {Patience}
  J., {Lafreni{\`e}re} D., {Doyon} R., 2008, Science, 322, 1348

\bibitem[{{Marois} {et~al}\mbox{.}(2010){Marois}, {Zuckerman}, {Konopacky},
  {Macintosh}, \& {Barman}}]{2010Natur.468.1080M}
{Marois} C., {Zuckerman} B., {Konopacky} Q.~M., {Macintosh} B., {Barman} T.,
  2010, \nat, 468, 1080

\bibitem[{{Masunaga} \& {Inutsuka}(2000)}]{2000ApJ...531..350M}
{Masunaga} H., {Inutsuka} S., 2000, \apj, 531, 350

\bibitem[{{Mellon} \& {Li}(2008)}]{2008ApJ...681.1356M}
{Mellon} R.~R., {Li} Z.-Y., 2008, \apj, 681, 1356

\bibitem[{{Nayakshin}(2010{\natexlab{a}})}]{2010MNRAS.408L..36N}
{Nayakshin} S., 2010{\natexlab{a}}, \mnras, 408, L36

\bibitem[{{Nayakshin}(2010{\natexlab{b}})}]{2010MNRAS.408.2381N}
---, 2010{\natexlab{b}}, \mnras, 408, 2381

\bibitem[{{Palla} \& {Stahler}(1993)}]{1993ApJ...418..414P}
{Palla} F., {Stahler} S.~W., 1993, \apj, 418, 414

\bibitem[{{Price}(2007)}]{2007PASA...24..159P}
{Price} D.~J., 2007, PASA, 24, 159

\bibitem[{{Rice}, {Lodato} \& {Armitage}(2005){Rice}, {Lodato}, \&
  {Armitage}}]{2005MNRAS.364L..56R}
{Rice} W.~K.~M., {Lodato} G., {Armitage} P.~J., 2005, \mnras, 364, L56

\bibitem[{{Seifried} {et~al}\mbox{.}(2013){Seifried}, {Banerjee}, {Pudritz}, \&
  {Klessen}}]{2013MNRAS.432.3320S}
{Seifried} D., {Banerjee} R., {Pudritz} R.~E., {Klessen} R.~S., 2013, \mnras,
  432, 3320

\bibitem[{{Semenov} {et~al}\mbox{.}(2003){Semenov}, {Henning}, {Helling},
  {Ilgner}, \& {Sedlmayr}}]{2003A&A...410..611S}
{Semenov} D., {Henning} T., {Helling} C., {Ilgner} M., {Sedlmayr} E., 2003,
  \aap, 410, 611

\bibitem[{{Stamatellos} {et~al}\mbox{.}(2011){Stamatellos}, {Maury},
  {Whitworth}, \& {Andr{\'e}}}]{2011MNRAS.413.1787S}
{Stamatellos} D., {Maury} A., {Whitworth} A., {Andr{\'e}} P., 2011, \mnras,
  413, 1787

\bibitem[{{Stamatellos} \&
  {Whitworth}(2009{\natexlab{a}})}]{2009MNRAS.392..413S}
{Stamatellos} D., {Whitworth} A.~P., 2009{\natexlab{a}}, \mnras, 392, 413

\bibitem[{{Stamatellos} \&
  {Whitworth}(2009{\natexlab{b}})}]{2009MNRAS.400.1563S}
---, 2009{\natexlab{b}}, \mnras, 400, 1563

\bibitem[{{Stamatellos}, {Whitworth} \& {Hubber}(2011){Stamatellos},
  {Whitworth}, \& {Hubber}}]{2011ApJ...730...32S}
{Stamatellos} D., {Whitworth} A.~P., {Hubber} D.~A., 2011, \apj, 730, 32

\bibitem[{{Takahashi}, {Inutsuka} \& {Machida}(2013){Takahashi}, {Inutsuka}, \&
  {Machida}}]{2013ApJ...770...71T}
{Takahashi} S.~Z., {Inutsuka} S., {Machida} M.~N., 2013, \apj, 770, 71

\bibitem[{{Tanaka}, {Takeuchi} \& {Ward}(2002){Tanaka}, {Takeuchi}, \&
  {Ward}}]{2002ApJ...565.1257T}
{Tanaka} H., {Takeuchi} T., {Ward} W.~R., 2002, \apj, 565, 1257

\bibitem[{{Thalmann} {et~al}\mbox{.}(2009){Thalmann}, {Carson}, {Janson},
  {Goto}, {McElwain}, {Egner}, {Feldt}, {Hashimoto}, {Hayano}, {Henning},
  {Hodapp}, {Kandori}, {Klahr}, {Kudo}, {Kusakabe}, {Mordasini}, {Morino},
  {Suto}, {Suzuki}, \& {Tamura}}]{2009ApJ...707L.123T}
{Thalmann} C. {et~al.}, 2009, \apjl, 707, L123

\bibitem[{{Tomida} {et~al}\mbox{.}(2013){Tomida}, {Tomisaka}, {Matsumoto},
  {Hori}, {Okuzumi}, {Machida}, \& {Saigo}}]{2013ApJ...763....6T}
{Tomida} K., {Tomisaka} K., {Matsumoto} T., {Hori} Y., {Okuzumi} S., {Machida}
  M.~N., {Saigo} K., 2013, \apj, 763, 6

\bibitem[{{Tsukamoto}, {Iwasaki} \& {Inutsuka}(2013){Tsukamoto}, {Iwasaki}, \&
  {Inutsuka}}]{2013MNRAS.434.2593T}
{Tsukamoto} Y., {Iwasaki} K., {Inutsuka} S., 2013, \mnras, 434, 2593

\bibitem[{{Tsukamoto} \& {Machida}(2011)}]{2011MNRAS.416..591T}
{Tsukamoto} Y., {Machida} M.~N., 2011, \mnras, 416, 591

\bibitem[{{Tsukamoto} \& {Machida}(2013)}]{2013MNRAS.428.1321T}
---, 2013, \mnras, 428, 1321

\bibitem[{{Vorobyov} \& {Basu}(2010)}]{2010ApJ...714L.133V}
{Vorobyov} E.~I., {Basu} S., 2010, \apjl, 714, L133

\bibitem[{{Vorobyov}, {DeSouza} \& {Basu}(2013){Vorobyov}, {DeSouza}, \&
  {Basu}}]{2013ApJ...768..131V}
{Vorobyov} E.~I., {DeSouza} A.~L., {Basu} S., 2013, \apj, 768, 131

\bibitem[{{Whitehouse} \& {Bate}(2004)}]{2004MNRAS.353.1078W}
{Whitehouse} S.~C., {Bate} M.~R., 2004, \mnras, 353, 1078

\bibitem[{{Whitehouse}, {Bate} \& {Monaghan}(2005){Whitehouse}, {Bate}, \&
  {Monaghan}}]{2005MNRAS.364.1367W}
{Whitehouse} S.~C., {Bate} M.~R., {Monaghan} J.~J., 2005, \mnras, 364, 1367

\bibitem[{{Zhu} {et~al}\mbox{.}(2012){Zhu}, {Hartmann}, {Nelson}, \&
  {Gammie}}]{2012ApJ...746..110Z}
{Zhu} Z., {Hartmann} L., {Nelson} R.~P., {Gammie} C.~F., 2012, \apj, 746, 110

\end{thebibliography}

\end{document}